\documentclass[11pt, aps, showkeys,showpacs]{revtex4}
\usepackage{graphicx}
\usepackage{amssymb}
\usepackage{amsmath}
\usepackage{latexsym}

\begin{document}

\title{The Localization of $s$-Wave and Quantum Effective Potential of a
Quasi-Free Particle with Position-Dependent Mass}

\author{Guo-Xing JU, Yang Xiang and Zhong-Zhou Ren}

\affiliation{Department of Physics, Nanjing University, Nanjing
210093,China}


\email[Corresponding author:]{jugx@nju.edu.cn}

\begin{abstract}
The properties of the $s$-wave for  a quasi-free particle with position-dependent
mass(PDM) have been discussed in details. Differed from
the system with constant mass in which the localization of the $s$-wave for the free quantum particle
around the origin  only occurs in two dimensions, the quasi-free particle with PDM
can experience  attractive forces in $D$ dimensions except $D=1$
when its mass function  satisfies some conditions.
The effective mass of a particle varying  with its position  can induce effective
interaction which may be attractive in some cases. The  analytical expressions of the eigenfunctions
and the corresponding probability densities for the $s$-waves of the two-
and three-dimensional systems with a special PDM  are given,
and the existences of localization around the origin for these systems are shown.
\end{abstract}



\maketitle


\section{Introduction}

The properties of a system interested in physics have a close relationship with its
dimensionality of space. Quantum Hall effect is a
macroscopical quantum phenomenon occurred in the two-dimensional
system\cite{pra}. Bose-Einstein condensation only takes place in the
three-dimensional Bose systems but not in one- and
two-dimensional cases\cite{huang}. There
is no finite-temperature phase transition in the one-dimensional system, and there may exist
Kosterlitz-Thouless transition of the topological type in two-dimensions\cite{kos-tho}
and the conventional thermodynamics phase transition in
three-dimensions\cite{huang}. One has made extensive theoretical and
experimental researches on  properties of the systems with various
dimensions and also developed many effective methods to understand them. So far,  quantum
mechanics is  believed to be the most successful theory to tackle various problems in physics.

In quantum mechanics, we usually solve Schr\"{o}dinger equations of various potentials to
obtain the eigenfunctions  and the related physical quantities of the system.
It is well known that the wave functions for both the
two-dimensional  system with axial symmetries and the three-dimensional system with
spherical symmetry  can be separated into the product of the radial and angular parts. Due to the fact that
the  angular component of the
eigenfunction is the spherical harmonics(in three-dimensions) or its reduced form(in two-dimensions), the
emphasis to solve Schr\"{o}dinger equation is put on its radial component for the system
with spherical or axial symmetries\cite{flu}.
The radial Schr\"{o}dinger  equations for two- and three-dimensional systems
have little difference in their forms, we can transform one  into another by just making a
simple replacement of the quantum number of angular momentum. It is in this sense that
there is no remarkable difference between two- and
three-dimensional systems in the framework of quantum mechanics as in other problems. However,
Cirone and  coworkers have made systematic studies on
the two-dimensional system through solving the  Schr\"{o}dinger
equation in recent
years\cite{ber,cir,bia,sch,bot,jen,bia2, cir2}. They find that the
$s$-state of a free quantum particle in two dimensions can have a unique localization around the origin
which does not present in other dimensions.

The study of quantum systems with position-dependent mass(PDM)
has become one of the active subjects over the last years. These systems have found
wide applications in the determination of electrical properties of various
systems such as semiconductor, quantum dot, liquid crystal and so
on. Theoretically, the search for the exact solutions of the
wave equations, especially those of the Schr\"{o}dinger equation
with PDM has aroused  great
interest\cite{von,bas,ser,bar,lev,plas,mil,de,go1,go2,alh,koc, que1,que2,que3,que4,
cai}, they may provide useful models to the description of the systems with PDM.
A natural problem is what relations there exist between the
dimensionality of space and  properties of the system in the
case of PDM. Up to now, the study of the system with PDM is
mainly focused on one rather than higher dimensions. As far as we know, there are no
investigations on the possible differences between systems of
different dimensions with PDM. In this article, we will
discuss the properties of systems of different dimensions with PDM
through their zero angular momentum states ($s$-state). We find that the
effective attractive potential appeared in the two-dimensional
system with constant mass also exists in the two-dimensional
system with PDM, and in particular that there is an additional
effective potential related to the mass function of the particle
in the latter case. The effective potential induced by the
variation of mass with position may be attractive or repulsive
depending on the sign of the first order derivative of the mass
function with respect to the position. The more important thing is
that this induced effective potential appears in the system with any dimension
other than one. When effective mass
is a constant, the induced effective potential
vanishes, which  reflects the fact that the system  with PDM is more
general than that of the constant mass. In this article, we will
give the explicit expression of the effective potential for the quasi-free particle
with PDM in $D$ dimensions. For the so-called quasi-free particle we mean that the particle does not
subject to other potential field except its mass is position-dependent. With the mass being constant,
the quasi-free particle
becomes free one.
We also solve the radial Schr\"odinger equation to
get the eigenfunctions of the $s$ states and the corresponding probability density
functions, and analyze their behaviors for a special mass function in the two- and three-dimensional
systems, respectively.

This article is organized as follows: In section \ref{rsch},  we will analyze the
$s$-state of the quasi-free particle with PDM and make a comparison between the properties of
the systems in different dimensionality $D$ as well as btween those of the system with constant mass
and with PDM in terms of the effective potential. In sections \ref{td2} and \ref{td3}, we
will get  the exact solutions for the $s$-wave  of the quasi-free particle with a
special mass function in two and three dimensions,
respectively. They  localize around the origin, which show
the existence of the effective attractive
potential. In Section \ref{con}, we will make some conclusions and discussions.

\section{Radial Schr\"{o}dinger equations and the effective potentials of $D$-dimensional quasi-free particle with PDM}
\label{rsch}

We first consider a $D$-dimensional system with PDM. We
will assume that the
effective mass  varies with the radial coordinate $m=m(r)$. When the mass of a
particle depends on its position, the mass and momentum operators
no longer commute, so there are several ways to define the kinetic
energy operator of the quantum system. In this paper, we will
adopt the form of the kinetic energy introduced by Levy-Leblond
and etc\cite{lev}, which reads
\begin{equation}\label{kn}
    T=\frac{1}{2m_0}\vec{\bf p}\frac{1}{\mu(r)}\cdot\vec{\bf p},
\end{equation}
where $\mu(r)=\frac{m(r)}{m_0}$ is dimensionless mass, and the momentum operator
$\vec{\bf p}$ can be expressed in terms of $D$-dimensional differential
operator $\nabla_D$ as
$$
\vec{\bf p}=-i\hbar\nabla_D.
$$
In Cartesian coordinate, we have the representation of the operator $\nabla_D$ to be of the form
$\nabla_D=(\frac{\partial }{\partial
x_1},\,\frac{\partial }{\partial x_2}, \,\cdots,\frac{\partial
}{\partial x_D})$. In the following sections, we will use the units of $m_0=1, \hbar=1$,
then the Schr\"{o}dinger equation of the quasi-free particle
with PDM in $D$-dimensional space may be written as
\begin{equation}
\label{e15}
\left( - \frac{1}{2}\vec \nabla _D \frac{1}{{\mu (r)}}\cdot\vec\nabla _D \right)\psi (\vec r) = E\psi (\vec r).
\end{equation}

We adopt spherical coordinate $(r,\theta_1,\theta_2,\cdots,
\theta_{D-2},\phi)$,
and define a operator
\begin{equation}
\label{e18}
T_r  = \frac{1}{2}\hat p_r \frac{1}{{\mu (r)}}\hat p_r,
\end{equation}
where
$$
\hat p_r =\frac{1}{2}(\hat{\vec{r}}\cdot\vec{p}+\vec{p}\cdot\hat{\vec{r}})
=- i\frac{1}{{r^{(D - 1)/2} }}\frac{\partial
}{{\partial r}}r^{(D - 1)/2}=-i\left(\frac{\partial}{\partial r}+\frac{D-1}{2r}\right),
$$
is the radial momentum operator in $D$-dimensional space\cite{nie, sch, ban, kos, ave,levk, paz}.
The operator $T_r$ reduces to
$\frac{{\hat p_r ^2 }}{{2\mu }}$ with the mass being constant.
We now rewrite the Hamiltonian of the quasi-free particle
with PDM in $D$ dimensions as follows\cite{ban,nie, kos,ave, levk}
\begin{equation}
\label{e19}
H = T = T_r  - \frac{1}{{2\mu (r)}}\left\{  - \frac{{\hat{L}^2 }}{{r^2 }} + \frac{{\mu '}}{\mu } \cdot \frac{{(D - 1)}}{{2r}}
- \frac{{(D - 1)(D - 3)}}{{4r^2 }}\right\},
\end{equation}
where $\hat{L}$ is the $D$-dimensional angular momentum operator.

For the $D$-dimensional   system with spherical symmetry, the eigenfunction
$\psi (\vec r)$ of the system can be made separation of variables\cite{ban,nie,kos,ave,levk}
\begin{equation}
\label{e16}
 \psi (\vec r) = r^{ - (D - 1)/2} R(r) \cdot Y_{l_{D - 2}, \cdots,l_1 }^l
 (\theta _1 ,\theta _2 , \cdots, \theta _{D - 2},\phi ),
\end{equation}
where $Y_{l_{D - 2} , \cdots, l_1 }^l (\theta _1 ,\theta _2 ,
\cdots,\theta _{D - 2},\phi )$ is $D$-dimensional hyperspherical harmonic
function, and $l$ is the quantum number of angular momentum.

With Eq.(\ref{e19}) and the eigenfunction (\ref{e16}), we
can get the $D$-dimensional radial Schr\"{o}dinger equation from Eq.(\ref{e15})
\begin{equation}
\label{e22}
\left[ - \frac{d}{{dr}}\frac{1}{{2\mu
(r)}}\frac{d}{{dr}} + \frac{1}{{2\mu (r)}}\frac{{l(l + D -
2)}}{{r^2 }} - \frac{{\mu '}}{{\mu ^2 }} \cdot \frac{{(D -
1)}}{{4r}} + \frac{{(D - 1)(D - 3)}}{{8\mu r^2 }}\right]R(r) =
ER(r),
\end{equation}
where $\frac{1}{{2\mu (r)}}\frac{{l(l + D - 2)}}{{r^2}}$ is the centrifugal
potential which results from the action of the term $\frac{1}{2\mu (r)} \frac{\hat{L}^2 }{r^2 }$
in Eq.(\ref{e19}) on the hyperspherical harmonic function in Eq.(\ref{e16}). The factor $l(l+D-2)$
is the eigenvalue of the operator $\hat{L}^2$ with eigenfunction being the hyperspherical
harmonic function\cite{ban, nie, ave}.
If we only consider $s$-wave, the contribution from the centrifugal potential
is zero. The last two terms in the bracket on LHS of Eq.(\ref{e22}) are the quantum effective
potential (QEP) for the system with PDM,
\begin{equation}
\label{e20}
V_{QEP} =  - \frac{{\mu '}}{{\mu ^2 }}\frac{{(D
- 1)}}{{4r}} + \frac{{(D - 1)(D - 3)}}{{8\mu r^2 }}.
\end{equation}
The first term $ - \frac{{\mu '}}{{2\mu ^2 }}  \frac{{(D - 1)}}{{2r}}$ in Eq.(\ref{e20})
is the effective potential due to the dependence of the mass on position;The
second term comes from the reduction of the motion of the particle in $D$ dimensions
to that of radial component and it also appears in the system with constant mass.

 Now we can make a few remarks on  Eq.(\ref{e20}): (i) A quasi-free particle
with PDM is generally  not free in the radial direction. When
$D=3$ and the mass of the particle is constant, QEP disappears. However, in the system with PDM,
the potential $-\frac{{\mu '}}{{2\mu ^2  r}}$ originated from the variation
of mass with position  does not vanish. For $D=1$, QEP
is zero for systems with constant mass or PDM. So QEP remains to be related to the dimensionality  of the system.
 (ii) For $D>3$, QEP is a repulsive potential if the mass of the particle decreases with the radial coordinate,
 i.e. $\mu'(r)<0$.
 (iii) Being different from the system with constant mass in which attractive QEP only occurs for  $D=2$,
 we may also have attractive QEP for other $D$ in the systems with PDM.
 The sign of QEP in
 Eq. (\ref{e20}) determines whether QEP is attractive or repulsive. For example, the QEP is attractive
 for $D=3$ when the mass function $\mu(r)$
 satisfies the condition $\mu'(r)>0$.


\section{Eigenfunctions and Localization of the $s$ States for a Quasi-Free Particle with PDM}

\subsection{Two-dimensional system}
\label{td2}

When $D=2$, the Hamiltonian
of a quasi-free particle with PDM in polar coordinate $(\rho, \varphi)$ can be written as follows from Eq.(\ref{e19})
\begin{equation}
\label{e8}
H=T=\frac{1}{2}\hat p_\rho  \frac{1}{\mu (\rho )}\hat
p_\rho   - \frac{1}{2\mu \rho ^2 }\frac{\partial ^2
}{\partial \varphi ^2 } - \frac{1}{8\mu (\rho )\rho ^2 } -
\frac{\mu '(\rho )}{4\mu ^2 (\rho )\rho },
\end{equation}
where $ \hat p_\rho = - i\hbar (\frac{\partial }{\partial \rho } +
\frac{1}{2\rho }) $ is the radial momentum operator for the
two-dimensional system.

The wave function of the two-dimensional system  with axial symmetry
can have the following form
\begin{equation}\label{twf}
    \psi(\vec{\rho})=\frac{1}{\sqrt{\rho}}R(\rho)e^{il\varphi},
\end{equation}
where $l=0, \pm1, \pm2, \cdots$ is the  quantum number of angular
momentum. If we only consider $s$-state, that is $l=0$ in
Eq.(\ref{twf}), then the  second term  in Eq.(\ref{e8}) will
disappear in the  radial Schr\"{o}dinger equation. The last two
terms in Eq. (\ref{e8}) is the QEP  mentioned-above and is the
reduction of Eq.(\ref{e20}) in two dimensions,
\begin{equation}
\label{e9} V_{QFP}  =  - \frac{1}{8\mu (\rho )\rho ^2 } -
\frac{\mu '(\rho )}{4\mu ^2 (\rho )\rho }.
\end{equation}
From Eq.(\ref{e9}), the following remarks are in order:
(i) When mass is a constant, i.e.
$\mu'=0$, $V_{QEP}$  reduced to that studied by Cirone and coworkers
\cite{ber, cir,bia,sch,bot,jen,bia2, cir2}. (ii) If the mass
$\mu$ increases with $r$, that is $\mu' >0$, then $V_{QFP}$ is always
negative, and  so it is a central attractive potential. In this case,
the variation of mass strengthens the attractive force. (iii) If $\mu'<0$, QEP
may be attractive or repulsive depending on the specific form of the mass function. However,
the dependence of mass on the position may weaken  and even destroy the
attractive action in this case. So in the two-dimensional system, a quasi-free particle
with PDM is subject to an effective potential in its radial motion. This effective potential
has an  additional part from the variation
of effective mass except for that occurred  in the system with constant mass.

Now we devote ourself to the solution of the $s$-state and discuss its properties
for a given mass function. We have known that if the effective mass $\mu$ increases with $r$,
i.e. $\mu' >0$, then QEP $V_{QFP}$ is always
attractive. When we only consider $s$-wave, the effect from the centrifugal potential is
suppressed, so  we expect that there
exists the bound state localized around the origin and the
corresponding probability function displays a maximum near the origin.

We chose the mass function to be of the form
$\mu(\rho)=\rho^\alpha$, and we also assume that the energy  be
negative, i.e. $E =-|E|$ so that the state is  bound. Notice that
the mass function may have a positive multiple factor, we group it
into $m_0$. The requirement that the mass function satisfies the
condition $\mu'>0$ imposes the restriction $\alpha>0$ on its
exponent $\alpha$. This is a stronger restriction for $\alpha$.
Because we only require that $V_{QFP}$ be attractive, that is
$V_{QFP}<0$, Eq. (\ref{e20}) implies that $\alpha>-\frac{1}{2}$ is
sufficient for that requirement with the above mass function. Due to
the following considerations, we need not worry about the
possibility of the above effective mass increasing infinitely: We
will soon  see that the eigenfunction decays exponentially with the
distance  from the origin and so the increment of the effective mass
is finite around the origin. On the other hand, we mainly concern
the behavior of the eigenfunction for the $s$ state around the
origin. This situation is similar to that  of  a particle under the
harmonic oscillator potential.

Now  the radial Schr\"{o}dinger equation for the quasi-free particle with
effective mass of the form $\mu(\rho)=\rho^\alpha$ can be written as
\begin{equation}
\label{e10}
 -\frac{1}{2}\frac{d^2 u(\rho)}{d \rho^2 }-\frac{1-\alpha}{2\rho}\frac{d u(\rho)}{d \rho }
+ \frac{l^2 }{2\rho^2 }u(\rho)  =  -|E|\rho^\alpha u(\rho),
\end{equation}
where $u(\rho)$ is the radial wave function which is related to
$R(\rho)$ in Eq.(\ref{twf})  through the equation
\begin{equation}
 u(\rho)=\frac{R(\rho)}{\sqrt{\rho}}.
\end{equation}
For $s$-wave, i.e. $l=0$, Eq. (\ref{e10}) reduces to the following equation
\begin{equation}
\label{e11} \frac{d^2 u }{d\rho ^2 }+\frac{1-\alpha}{\rho}\frac{d
u(\rho)}{d\rho }-2|E|\rho^{\alpha} u(\rho)   = 0.
\end{equation}
Setting
\begin{equation}\label{para}
    z=\frac{2\sqrt{2|E|}}{\alpha+2}\rho^{(\alpha+2)/2},\;
    \nu=\frac{\alpha}{2+\alpha},\;
    u(\rho)=z^\nu \xi(z),
\end{equation}
we may transform Eq.(\ref{e11}) into the standard modified Bessel equation\cite{mag}
\begin{equation}\label{beeq}
z^2\frac{d^2 \xi(z) }{d z^2 }+z\frac{d \xi(z)}{dz
}-(z^2+\nu^2)\xi(z)= 0.
\end{equation}
Eq.(\ref{beeq}) has two special solutions $I_\nu(z)$ and
$K_{\nu}(z)$ which are the first and second modified Bessel
function, respectively. It should be noted  that the range of $\nu$ is $-\frac{1}{3}<\nu<1$
due to $\alpha>-\frac{1}{2}$.

\begin{figure}[htb]
  \centering
  \begin{minipage}[c]{0.5\textwidth}
    \centering
    \includegraphics[width=6.5cm,scale=0.4]{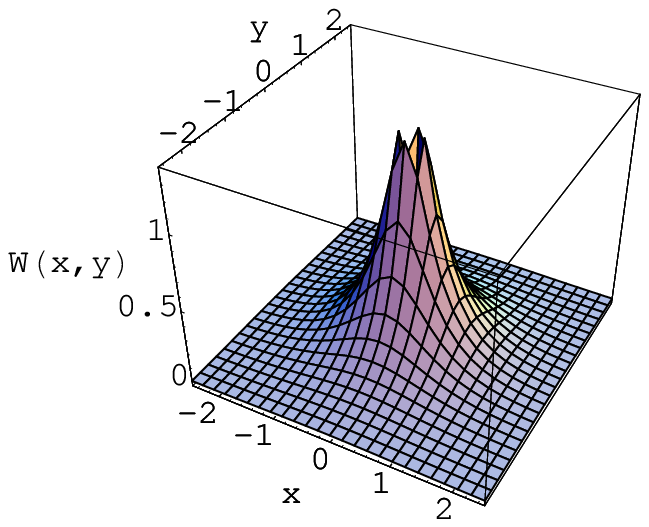}\\
    {(a)$\;\nu=-0.2$}
  \end{minipage}%
  \begin{minipage}[c]{0.5\textwidth}
    \centering
    \includegraphics[width=6.5cm,scale=0.4]{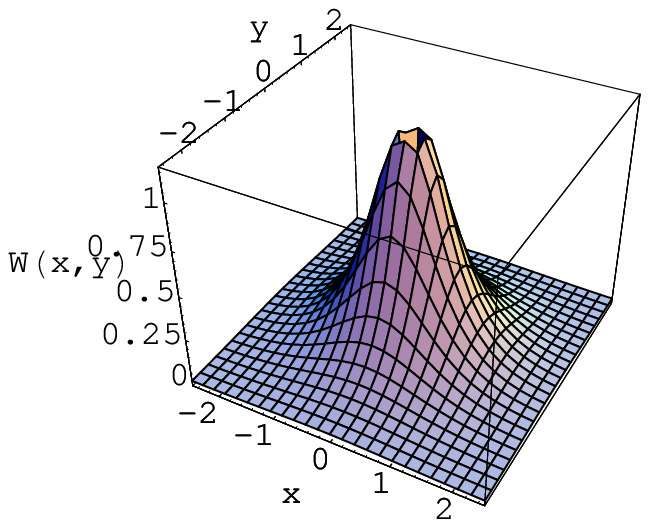}\\
    {(b)$\;\nu=0.1$}
  \end{minipage}
  \vspace*{0.5cm}
  \begin{minipage}[c]{0.5\textwidth}
    \centering
    \includegraphics[width=6.5cm,scale=0.4]{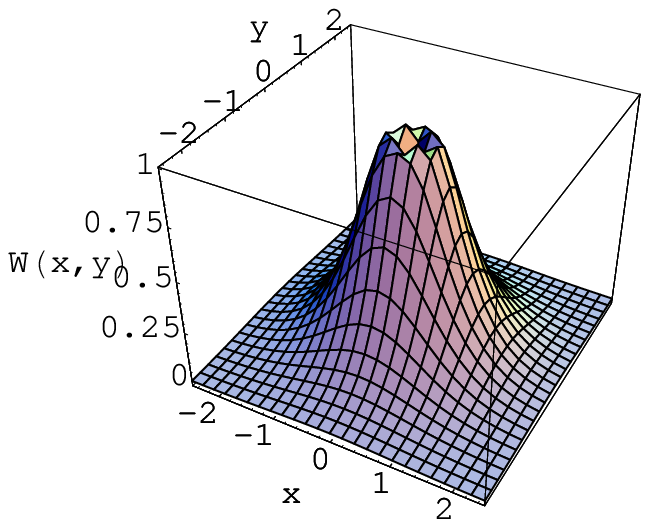}\\
    {(c)$\;\nu=0.4$}
  \end{minipage}%
  \begin{minipage}[c]{0.5\textwidth}
    \centering
    \includegraphics[width=6.5cm,scale=0.4]{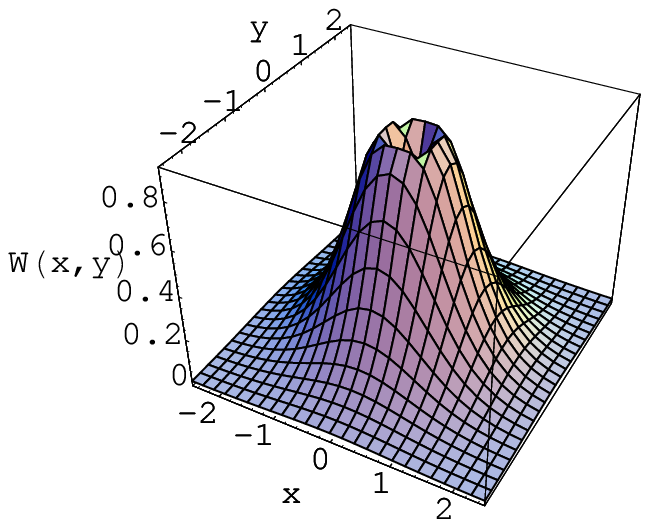}\\
    {(d)$\;\nu=0.7$}
  \end{minipage}
\caption{The radial probability density of $s$-state of a
two-dimensional quasi-free particle with PDM  is
$0$ at the origin, but has a maximum close to the origin. The radial probability density
decays exponentially for large $\rho$. The parameters are $|E|= \frac{1}{2}$,
and $\nu=-0.2,\, 0.1,\, 0.4,\, 0.7$, respectively.}
  \label{f1wf}
\end{figure}

\begin{figure}[htb]
  \centering
  \begin{minipage}[c]{0.5\textwidth}
    \centering
    \includegraphics[width=6.0cm]{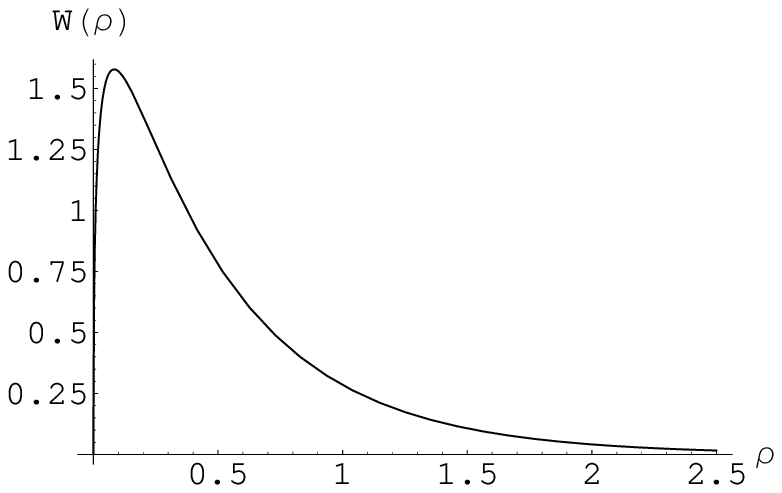}\\
    {(a)$\;\nu=-0.2$}
  \end{minipage}%
  \begin{minipage}[c]{0.5\textwidth}
    \centering
    \includegraphics[width=6.0cm]{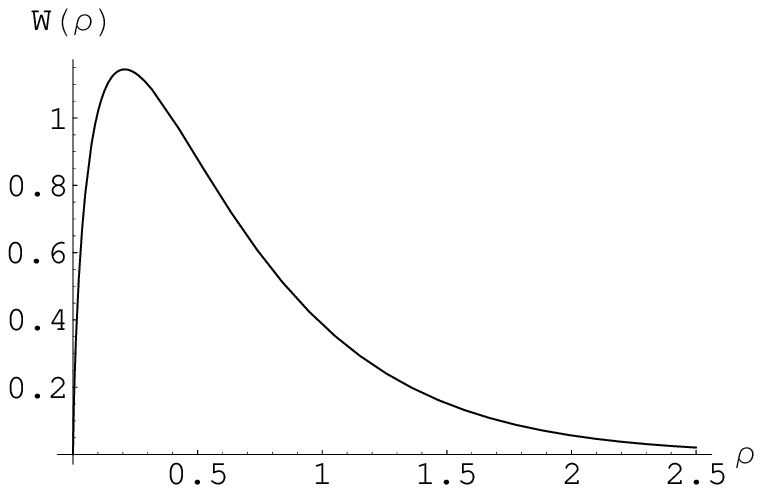}\\
    {(b)$\;\nu=0.1$}
  \end{minipage}
  \vspace*{0.5cm}
  \begin{minipage}[c]{0.5\textwidth}
    \centering
    \includegraphics[width=6.0cm]{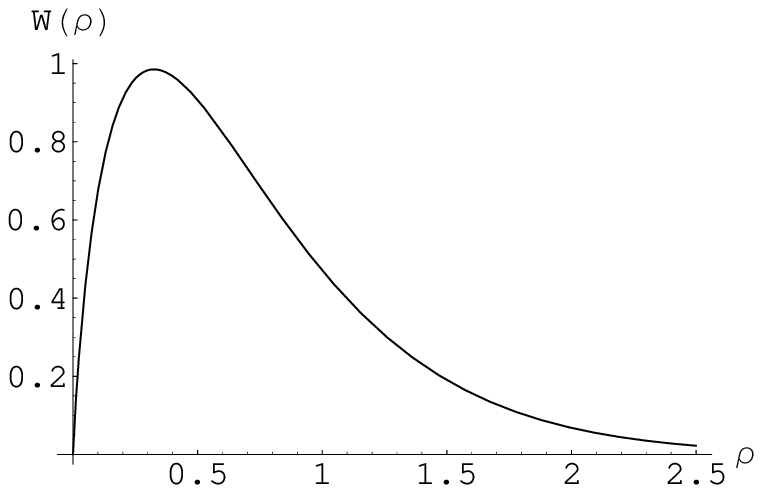}\\
    {(c)$\;\nu=0.4$}
  \end{minipage}%
  \begin{minipage}[c]{0.5\textwidth}
    \centering
    \includegraphics[width=6.0cm]{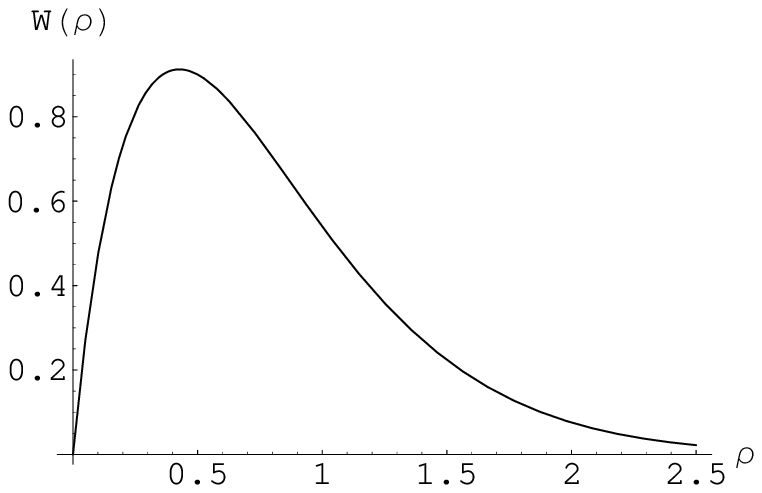}\\
    {(d)$\;\nu=0.7$}
  \end{minipage}
\caption{The cut of the radial probability density of the $s$-state for a
two-dimensional quasi-free particle with PDM  shows that the maximum of the probability
density deviates from he origin with $\nu$  increasing.  The parameters are the same as those in
Fig.\ref{f1wf}.}
 \label{f1pd}
\end{figure}

We require that $u$  be finite when $\rho$ is large so the eigenfunction may be normalized. This is
equivalent to $\xi(z)\rightarrow 0$ when $z\rightarrow\infty$.
Considering  of the asymptotic behavior of the functions $I_\nu$ and $K_\nu$ with
the modulus of their arguments being large\cite{mag}, we have the general solution of Eq.(\ref{beeq})
which is finite with large $|z|$,
\begin{equation}\label{besol}
    \xi=cK_\nu(z),
\end{equation}
where $c$ is a arbitrary constant. So the solution of Eq.(\ref{e11}) is
\begin{equation}
\label{e12}
\begin{aligned}
u(\rho) &=cz^{\nu}K_\nu(z)=\sqrt{\frac{\sin(\nu\pi)}{\nu\pi^2}}
\left[2|E|(1-\nu)\right]^{\frac{1}{2}(1-\nu)}z^{\nu}K_\nu(z) \\
&= \sqrt
{\frac{2|E|\sin(\nu\pi)(1-\nu)}{\nu\pi^2}}\rho^{\frac{\nu}{1-\nu}}
K_{\nu} ((1-\nu)\sqrt {2|E|} \rho ^{\frac{1}{1-\nu}} ).
\end{aligned}
\end{equation}
We can show that this eigenfunction has been normalized
\begin{equation}
\label{e13}
\begin{aligned}
\int_{-\infty}^{+\infty} dx \int_{-\infty}^{+\infty} dy
|\psi(x,y)|^{2} &=\int_{0}^{2\pi} d\varphi \int_{0}^{+\infty}
\rho\,d\rho |u(\rho)|^{2} \\
&=\frac{2\sin(\nu\pi)}{\nu\pi}\int_{0}^{\infty} z K^2_{\nu}(z)\,dz
=1.
\end{aligned}
\end{equation}
The probability density that the particle may appear between
$\rho$ and $\rho+d\rho$ is
\begin{equation}
\label{e14}
\begin{aligned}
W(\rho )d\rho  =& \int_0^{2\pi}d\theta\,\rho\,d\rho|\psi|^2
=2\pi|u(\rho)|^2\rho\,d\rho  \\
=&\frac{4|E|\sin(\nu\pi)(1-\nu)}{\nu\pi}\rho^{\frac{1+\nu}{1-\nu}}
K_{\nu}^2((1-\nu)\sqrt {2|E|} \rho ^{\frac{1}{1-\nu}} )\,d\rho.
\end{aligned}
\end{equation}
In Fig.\ref{f1wf}, we depict the probability density function $W(\rho)$ with
$|E|=\frac{1}{2}$ and $\nu=-0.2,0.1,0.4,0.7$, respectively. These
figures indicate that $W(\rho)$ sharply collapses at the origin
and decays exponentially for large $\rho$, with a maximum close to the origin.
In addition, the maximum of $W$ deviates from the origin as increasing  $\nu$.
In order to show how $W(\rho)$ varies with $\rho$, we  give the cuts of
Fig.\ref{f1wf} along the radial in  Fig.\ref{f1pd}. It is seen that the variation of $W(\rho)$ with $\rho$ is
similar to that  with constant mass as given in \cite{cir}. For other energies $E$,  the corresponding $W(\rho)$
has similar behavior, they all have the localization around the origin.

The localization of the probability density for the $s$ state of a quasi-free particle with PDM
is produced by above QEP $V_{QFP}$. Because the mass function is chosen so that $\mu'$ is
always positive,  $V_{QFP}$ is an attractive potential. The above localization of
the  probability density confirms that the  potential is attractive.

If $E=|E|>0$, we make the similar transformation as in Eq. (\ref{para}),
then the radial Schr\"odinger equation for $l=0$ and the mass function
$\mu(\rho)=\rho^{\alpha}$ reads
\begin{equation}\label{be-po}
z^2\frac{d^2 \xi(z) }{d z^2 }+z\frac{d \xi(z)}{dz
}+(z^2-\nu^2)\xi(z)= 0.
\end{equation}
Eq.(\ref{be-po}) has two special solutions $J_\nu(z)$ and $Y_\nu(z)$ which are first and second kind of
Bessel functions, respectively\cite{mag}. Considering of the asymptotic behavior of $J_\nu(z)$
and $Y_\nu(z)$ around the origin and large $|z|$
we have the  solution of Eq.(\ref{be-po}) to be of
the form
\begin{equation}\label{e-po}
    \xi=CJ_\nu(z)+DY_\nu(z),
\end{equation}
where $C$ and $D$ are constants. The radial eigenfunction for the positive energy is
\begin{equation}\label{rs-po}
    u(\rho)=z^{\nu}(CJ_\nu(z)+DY_\nu(z)).
\end{equation}
The related probability density function of the above eigenfunction does not possess of the property of localization
near the origin.

\subsection{Three-dimensional system}
\label{td3}

We use the same form of mass function  $\mu(r)=r^\alpha$ as that for the two-dimensional system, then the radial
Schr\"{o}dinger equation for the three-dimensional quasi-free particle is
\begin{equation}
\label{e10t}
 -\frac{1}{2}\frac{d^2 u(r)}{d r^2 }+\frac{1}{2}\frac{2-\alpha}{r} \frac{d u(r)}{dr }
+ \frac{l(l+1) }{2 r^2 }u(r)=-|E| r^\alpha u(r),
\end{equation}
where $u(r)$ is the radial wave function which has  the relation to
$R(r)$ in Eq.(\ref{e16})
\begin{equation}
 u(r)=\frac{R(r)}{r}.
\end{equation}
For the state of $l=0$, Eq.(\ref{e10t}) reads
\begin{equation}
\label{e11t} \frac{d^2 u(r)}{d r^2 }+\frac{2-\alpha}{r} \frac{d
u(r)}{dr } =2|E| r^\alpha u(r).
\end{equation}
Making the replacements of the  following forms
\begin{equation}\label{parat}
    z=\frac{2\sqrt{2|E|}}{\alpha+2}r^{(\alpha+2)/2},\;
    \nu=\frac{\alpha-1}{2+\alpha},\;
    u(\rho)=z^\nu \xi(z),
\end{equation}
we will  turn  Eq.(\ref{e11t}) into the standard modified
Bessel equation as Eq.(\ref{beeq}). Now, $\nu$ is restricted in  the range
$(-\frac{1}{2},\,1)$ due to $\alpha>0$.

Similar to the analysis of the two-dimensional
system and with the requirement that $u$ is finite when $r$ is large, the
general solution of Eq.(\ref{beeq}) is
\begin{equation}\label{besolt}
    \xi=cK_\nu(z),
\end{equation}
where $c$ is an arbitrary constant. The normalized solution of Eq.(\ref{e11t}) is
\begin{equation}
\label{e12t}
\begin{aligned}
u(r) &=cz^{\nu}K_\nu(z)=\sqrt{\frac{\sin(\nu\pi)}{2\nu\pi^2}}
\left[\frac{2}{3}(1-\nu)\right]^{\frac{1}{2}(1-2\nu)}\left(2|E|\right)^{\frac{1}{2}(1-\nu)}z^{\nu}K_\nu(z) \\
&= \sqrt
{\frac{8|E|\sin(\nu\pi)(1-\nu)}{3\nu\pi}}r^{\frac{\nu}{2(1-\nu)}}
K_{\nu} (\frac{2}{3}(1-\nu)\sqrt {2|E|} r ^{\frac{3}{2(1-\nu)}} ).
\end{aligned}
\end{equation}
The probability of finding the particle in the range $r$ to $r+dr$ is
\begin{equation}
\label{e14t}
\begin{aligned}
W(r )\,dr  =& \int_0^{2\pi}d\theta\int_0^{\pi}\sin\varphi\,d\varphi\,r^2\,dr|\psi|^2 \\
=&\int_0^{2\pi}d\theta\int_0^{\pi}\sin\varphi\,d\varphi\,r^2\,dr|u(r)|^2|Y_{00}(\theta,\varphi)|^2
=|u(r)|^2r^2\,dr  \\
=&\frac{8|E|\sin(\nu\pi)(1-\nu)}{3\nu\pi}r^{\frac{2-\nu}{1-\nu}}
K_{\nu}^2(\frac{2}{3}(1-\nu)\sqrt {2|E|} r ^{\frac{3}{2(1-\nu)}}
)\,dr.
\end{aligned}
\end{equation}
For the three-dimensional system, the relationship between the probability
density $W(r)$ and $r$ is similar to that for the two-dimensional system.
There also exists localization of the wave function for the $s$ state around the origin.

Similar to the discussion for the two-dimensional system in the last subsection,
the eigenfunction and the related probability density function for the positive energy in three dimensions
have also no localization close the origin.

\section{Conclusions and Discussions}
\label{con}

In this article, we discussed the properties of the $s$-state of a quasi-free
particle with PDM. When mass varies with radial coordinate $r$, it induces additional
effective potential which may produce
central attractive force and may cause the wave function of the $s$-state
to localize around the origin. In $D$, except $D=1$, dimensional
system with the mass of the particle being position-dependent, the localization  appears, which
contrasts sharply with the system of constant mass. We solve the radial Schr\"odinger equation
to get the exact $s$-wave function for a specific mass function
in two and three dimensions, respectively. Their corresponding probability densities and the
related asymptotic behaviors  are also analyzed, and the results are consistent with
the general discussions.

There are many kinds of mass functions that are used in the investigation of the systems with PDM, but
it is difficult to get the exact solutions of the radial Schr\"odinger equations for these mass functions
in higher dimensional systems. So some numerical studies are required for the understanding of the properties
of the systems with these mass functions.

\section*{Acknowledgments}

The program is supported by National Natural Science Found for
Distinguished Young Scientists(10125521), Doctoral Fund of
Ministry of Education of China(20010284036), Major State Basic
Research Development Program(G2000077400), Knowledge Innovation
Project of Chinese Academy of Sciences(KJCX2-SW-N02), and National
Natural Science Found(60371013)


\end{document}